\documentclass[prl,reprint,showpacs,groupedaddress]{revtex4-1}
\usepackage{amsmath,amssymb,amsfonts}
\usepackage{graphicx}
\usepackage{txfonts}
\newcommand{\eqnrefp}[1]{{[Eq.~(\ref{#1})]}}

\newcommand{\eqnreft}[1]{{Eq.~(\ref{#1})}}
\newcommand{\eqnsreft}[2]{{Eqs.~(\ref{#1}) and (\ref{#2})}}
\newcommand{\figreft}[2]{Fig.~\ref{#1}#2}
\newcommand{\figsreft}[4]{Figs.~\ref{#1}#2 and \ref{#3}#4}
\newcommand{\figreftfull}[2]{Figure~\ref{#1}#2}
\newcommand{\figrefp}[2]{[Fig.~\ref{#1}#2]}

\newcommand{\x}{\theta}
\newcommand{\fo}{\hat{\Psi}(\x)}
\newcommand{\foh}{\hat{\Psi}^\dagger(\x)}
\newcommand{\g}{C}
\newcommand{\avg}[1]{\langle #1 \rangle}
\newcommand{\twovec}[2]{\left( \begin{array}{c} #1 \\ #2 \end{array} \right)}
\newcommand{\twomatrix}[4]{\left( \begin{array}{cc} #1 & #2 \\ #3 & #4 \end{array} \right)}
\newcommand{\is}{g_{T}}

\begin{document}

\title{Coherence and Instability in a Driven Bose-Einstein Condensate: A Fully Dynamical Number-Conserving Approach}

\author{T. P. Billam}
\email{t.p.billam@durham.ac.uk}
\author{S. A. Gardiner}
\affiliation{Department of Physics, Durham University, Durham DH1 3LE, United Kingdom}

\date{\today}

\begin{abstract}
We consider a Bose-Einstein condensate
driven by periodic $\delta$-kicks. In contrast to first-order descriptions,
which predict rapid, unbounded growth of the noncondensate in
resonant parameter regimes, the consistent treatment of condensate depletion in
our fully-time-dependent, second-order description acts to damp this growth,
leading to oscillations in the (non)condensate population and the coherence of
the system.
\end{abstract}


\maketitle


Central to the concept of a weakly interacting atomic Bose-Einstein condensate
(BEC) is that each of the $N$ component atoms can be considered to be in
approximately the same motional state; this is manifest through the mean-field
description of zero-temperature ($T=0$) BEC dynamics with the Gross-Pitaevskii
equation (GPE), which takes the form of a single-particle Schr\"{o}dinger wave
equation with an additional cubic nonlinear term
\cite{pitaevskii_stringari_2003}.  At finite temperature there is thermal
depletion of the condensate, which can be theoretically accounted for in a
variety of ways \cite{blakie_etal_ap_2008}.  Even at $T=0$ in a system of
finite size there is always a finite noncondensate fraction
\cite{pitaevskii_stringari_2003,castin_dum_pra_1998}, and one expects dynamics
within the BEC to cause significant particle transfer from the condensate to
the noncondensate fraction under quite general circumstances
\cite{castin_dum_prl_1997,gardiner_jmo_2002,gardiner_etal_pra_2000,zhang_etal_prl_2004}.
When rapid, such dynamical depletion has commonly been supposed to presage
destruction of the BEC as a coherent entity, however previous studies have been
hampered by the absence of a self-consistent treatment
\cite{gardiner_etal_pra_2000,zhang_etal_prl_2004}.  Applying the
number-conserving approach of Gardiner and Morgan
\cite{gardiner_morgan_pra_2007}, fully dynamically to second order, we have
carried out the first such self-consistent treatment of a specific example
system.

Our chosen test-system is based on the quantum $\delta$-kicked rotor
\cite{casati_etal_1979,moore_etal_prl_1995,ryu_etal_prl_2006,saunders_etal_pra_2007},
a paradigm quantum-chaotic system in which periodic driving leads to complex
behavior, including dynamical localization
\cite{casati_etal_1979,moore_etal_prl_1995} and quantum resonances (associated
with ballistic increase in the kinetic energy
\cite{casati_etal_1979,ryu_etal_prl_2006,saunders_etal_pra_2007}). Atom-optical
realizations of such systems
\cite{moore_etal_prl_1995,ryu_etal_prl_2006,oberthaler_etal_prl_1999} comprise
an exciting area of research into quantum-chaotic phenomena; extension into the
regime of BECs has also become an active area of research, in which several new
phenomena have been predicted
\cite{shepelyansky_prl_1993,mieck_graham_jpa_2004,monteiro_etal_prl_2009,zhang_etal_prl_2004,gardiner_etal_pra_2000}.
In the mean-field approximation, the GPE nonlinearity can strongly influence
$\delta$-kicked-rotor-BEC dynamics \cite{shepelyansky_prl_1993}; in particular,
the structure of quantum resonances was recently elucidated
\cite{monteiro_etal_prl_2009}, revealing previously unobserved resonance
profiles with a sharp asymmetric cut-off. The noncondensate fraction can be accounted for using a number-conserving approach \cite{castin_dum_pra_1998,castin_dum_prl_1997,gardiner_morgan_pra_2007,gardiner_pra_1997}: such approaches contain a fixed atom number (appropriate to atomic BEC experiments \cite{gardiner_morgan_pra_2007}), and nonlocal terms [$\tilde{f}$ and $Q$ in \eqnsreft{eqn_ggpe}{eqn_bdge}] which, within the second-order, linear-response treatment used by Morgan {\it et al.} \cite{morgan_etal_prl_2003}, have proved vital in explaining the temperature dependence of collective excitation spectra observed at JILA and MIT \cite{jin_etal_prl_1996}. The first-order, number-conserving description of
Castin and Dum \cite{castin_dum_prl_1997} [consisting of modified Bogoliubov-de
Gennes equations (MBdGE) coupled to the GPE] has been the approach of choice in $\delta$-kicked BEC
systems \cite{gardiner_etal_pra_2000,zhang_etal_prl_2004}. This has revealed a
general tendency towards rapid growth in the number of noncondensate atoms 
$N_{t}$.  Such behavior is directly linked to linear dynamical instabilities
(i.e., sensitivity to initial conditions in the linearized regime) in the GPE,
enabled by its nonlinearity \cite{gardiner_jmo_2002}; the presence of such
instabilities is a generic feature of most nonlinear systems.  Conversely,
growth in $N_{t}$ should match depletion in the condensate number $N_{c}$, and
as atoms transfer from $N_{c}$ to $N_{t}$, qualitatively one expects mean-field
interactions and hence further transfer to ``switch off'' at some stage.
Whether this occurs before the destruction of the condensate has remained an
open question, as a linearized, first-order description treats the condensate
as an effectively undepletable ``particle bath,'' feeling no effect from the
noncondensate.

\begin{figure}
\includegraphics[width=\columnwidth]{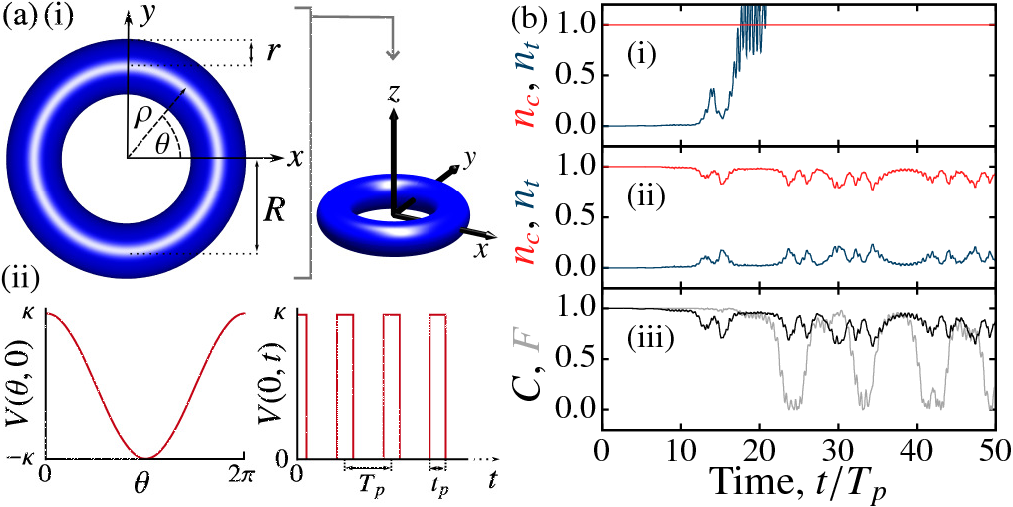}
\caption{(Color online) (a) $\delta$-kicked-rotor-BEC: (i) toroidally-trapped
BEC; (ii) $\delta$-kicking potential. (b) Evolution of condensate and
noncondensate fractions $n_c=N_c/N$ and $n_t=N_t/N$ in (i) first- and
(ii) second-order number-conserving descriptions ($N=10^4$,
$g_T=2.5\times10^{-4}$, $T_p=9.255$, $\kappa=0.5$). In (iii) we show the
coherence measure $C$ \eqnrefp{eqn_g_def} in the second-order description, and
the fidelity, $F$ \eqnrefp{eqn_f_def}, of the condensate mode between descriptions.}
\label{fig_system_results}
\end{figure}

In this Letter we answer this question in the affirmative, using the first fully time-dependent application of the second-order, number-conserving description of
Gardiner and Morgan \cite{gardiner_morgan_pra_2007}, for a
$\delta$-kicked-rotor-BEC in a quasi-one-dimensional ring trap
\figrefp{fig_system_results}{(a)}, an ideal test system to explore generic
issues of dynamically induced condensate depletion. This self-consistent description consists of a {\it
generalized} GPE (GGPE) coupled to the MBdGE; it explicitly preserves $N =
N_{c}+N_{t}$ and the orthogonality of the condensate and noncondensate,
includes mutual interactions, and allows free transfer of population between
the two. We solve these equations numerically and explore resonant parameter regimes which, to first order,
lead to rapid, unbounded growth of the noncondensate. Our
principal finding is the damping of this growth in the second-order description
\figrefp{fig_system_results}{(b)(i,ii)}. We also compute the coherence of the
system and the departure of the second-order description from the GPE
\figrefp{fig_system_results}{(b)(iii)} for varying $N$.  We show that, despite
considerable differences in dynamics between the descriptions around resonant
parameter regimes, the GPE accurately predicts the location of these parameter
regimes; the cut-off identified in \cite{monteiro_etal_prl_2009} is qualitatively preserved,
however the accompanying exponential oscillations are strongly modified for
experimentally realistic atom numbers.


We consider $N$ bosonic atoms of mass $M$, held in a toroidal potential
$V_{T}(\rho,z) = M\omega^2 [(\rho-R)^2+z^2]/2$
\figrefp{fig_system_results}{(a)(i)}, interacting with $s$-wave contact
interactions, and subject to a temporally and spatially periodic driving
potential $V(\theta,t)$ \figrefp{fig_system_results}{(a)(ii)}; toroidal
potentials similar to $V_T$ can be created and precisely  controlled using
all-optical methods \cite{ramanathan_etal_prl_2011}. Assuming sufficiently
strong radial and axial confinement, and harmonic length $r \equiv
\sqrt{\hbar/M\omega} \ll R$ \cite{halkyard_etal_pra_2010}, we reduce the system
Hamiltonian to a dimensionless (length unit $R$, time unit $MR^2/\hbar$),
one-dimensional form \cite{monteiro_etal_prl_2009}:
\begin{equation}
\hat{H} = \int d\x \foh \left[-\frac{1}{2}\frac{\partial^2}{\partial \x^2} 
+ V(\theta,t) + \frac{\is}{2} \foh \fo \right] \fo\,,
\label{eqn_many_body_hamiltonian}
\end{equation}
where the field operators obey bosonic commutation relations
$[\hat{\Psi}(\theta),\hat{\Psi}^{\dagger}(\theta')] = \delta(\theta-\theta')$.
The 
interaction strength $\is=2a_s R/r^2$, where $a_s$ is the
$s$-wave scattering length.  As in 
\cite{monteiro_etal_prl_2009,zhang_etal_prl_2004}, we model the driving
potential as a train of $\delta$-kicks, $V(\x,t) = \kappa \cos(\x)\sum_{n=
0}^{\infty} \delta(t-nT_p)$, with (dimensionless) kicking period $T_p$. This
may be approximated in experiment using short pulses of off-resonant laser
light \cite{moore_etal_prl_1995,ryu_etal_prl_2006,saunders_etal_pra_2007,Note1}.


We define the condensate mode $\psi(\x)$ (with 
creation operator
$\hat{a}_{c}^{\dagger}$) as the eigenfunction of the single-body density matrix
$\langle \hat{\Psi}^{\dagger}(\x')\hat{\Psi}(\x)\rangle$ with the largest
eigenvalue $N_{c}$ (the number of condensate atoms), to which it is normalized,
i.e., $\int d\x |\psi(\x)|^2 = N_c\equiv\langle
\hat{a}_{c}^{\dagger}\hat{a}_{c}\rangle$.  Hence, we expand the field operator as $\hat{\Psi}(\theta) = \hat{a}_{c}\psi(\x)/\sqrt{N_{c}} +
\delta\hat{\Psi}(\x)$, where $\delta\hat{\Psi}(\x) \equiv \int d\x Q(\x,\x')
\hat{\Psi}(\x')$ describes the field component orthogonal to the
condensate, and the projector $Q(\x,\x') = \delta(\x-\x') -
\psi(\x)\psi^{*}(\x')/N_{c}$.  We propagate $\psi(\x)$ with the GGPE:
\begin{equation}
\begin{split}
i \frac{\partial \psi(\x)}{\partial t} = & \left\{ H_{\textrm{GP}}(\x) - \lambda_{2} + \is \left[ 2\tilde{n}(\x,\x) - \frac{|\psi(\x)|^2}{N_c} \right] \right\} \psi(\x) \\
& + \is \tilde{m}(\x,\x) \psi^\ast(\x) - \is \tilde{f}(\x),
\end{split}
\label{eqn_ggpe}
\end{equation}
where $\lambda_{2}$ is a real-valued scalar constant with a role similar to the
chemical potential in grand-canonical treatments \cite{Note2}, and
\begin{equation}
H_{\textrm{GP}}(\x) \equiv  -\frac{1}{2}\frac{\partial^2}{\partial \x^2} + V(\x,t) + \is|\psi(\x)|^2.
\end{equation}
Introducing the number-conserving fluctuation operator $\tilde{\Lambda}(\theta)
\equiv \hat{a}_{c}^{\dagger} \delta\hat{\Psi}(\theta) / \sqrt{N_{c}}$
\cite{gardiner_morgan_pra_2007}, we define $\tilde{n}(\x,\x') \equiv \langle
\tilde{\Lambda}^{\dagger}(\x') \tilde{\Lambda} (\x) \rangle$ and
$\tilde{m}(\x,\x') \equiv \langle \tilde{\Lambda}(\x') \tilde{\Lambda}(\x)
\rangle$ (the noncondensate normal and anomalous averages), and
\begin{equation}
\tilde{f}(\x) \equiv \frac{1}{N_{c}}\int d\x' |\psi(\x')|^{2}\left[ \tilde{n}(\x,\x')\psi(\x') + \psi^{*}(\x')\tilde{m}(\x',\x)\right], 
\end{equation}
which ensures orthogonality of the condensate from the noncondensate
component.  The dynamics of $\psi(\x)$ are therefore coupled to those of
$\tilde{\Lambda}(\x)$, $\tilde{\Lambda}^{\dagger}(\x)$. We decompose
$\tilde{\Lambda}(\x) = \sum_{k=1}^{\infty} [\tilde{b}_{k}u_{k}(\x)
+\tilde{b}_{-k}u_{-k}(\x) + \tilde{b}_{k}^{\dagger}v_{k}^{*}(\x)+
\tilde{b}_{-k}^{\dagger}v_{-k}^{*}(\x)]$, where $\tilde{b}_{k},
\tilde{b}_{k}^{\dagger}$ are bosonic quasiparticle operators, and the
quasiparticle modes are normalized to $\int d\x [|u_{k}(\x)|^{2} -
|v_{k}(\x)|^{2}] = 1$, and choose all time-dependence to be within the
quasiparticle mode functions.  These are then propagated by the MBdGE:
\begin{equation}
i\frac{\partial}{\partial t} \twovec{u_k(\x)}{v_k(\x)} = \int d\x' \twomatrix{L(\x,\x')}{M(\x,\x')}{-M^\ast(\x,\x')}{-L^\ast(\x,\x')}  \twovec{u_k(\x')}{v_k(\x')},
\label{eqn_bdge}
\end{equation}
where
$
L(\x,\x') =  \delta(\x-\x')[H_{\textrm{GP}}(\x') - \lambda_{0}] 
+ \is \int d \x''Q(\x,\x'')|\psi(\x'')|^2 Q(\x'',\x')$, 
$M(\x,\x') = \is \int d \x''Q(\x,\x'')\psi(\x'')^2 Q^{*}(\x'',\x')$,
and $\lambda_{0}$  [the system ground state value of $(1/N_c)\int d\x  \psi^{*}(\x)
H_{\textrm{GP}}(\x) \psi(\x)$] is a lower-order approximant to $\lambda_{2}$.
At $T=0$, we can thus at all times express
$
\tilde{n}(\x,\x')  = \sum_{k=1}^{\infty} [v_k(\x)v_{k}^{*}(\x')+v_{-k}(\x)v_{-k}^{*}(\x')]
$,
$\tilde{m}(\x,\x')  = \sum_{k=1}^{\infty} [u_k(\x) v_k^\ast(\x')+u_{-k}(\x) v_{-k}^\ast(\x')]
$. The $k$ index quantifies the momentum 
associated with the equilibrium
quasiparticle eigenmodes, prior to application of 
$V(\x,t)$.
The coupling of \eqnreft{eqn_ggpe} and \eqnreft{eqn_bdge} constitutes the
second-order, number-conserving description of Gardiner and Morgan
\cite{gardiner_morgan_pra_2007}, and represents the equation of motion for $\hat{\Psi}(\x)$
expanded to second order in the fluctuation operator $\tilde{\Lambda}(\x)$ \cite{Note3}. 
To first-order
\cite{castin_dum_prl_1997,gardiner_pra_1997} one obtains the GPE
$i\partial \psi(\x)/\partial t = [H_{\textrm{GP}}(\x)-\lambda_{0}]\psi(\x)$,
coupled to the MBdGE \eqnrefp{eqn_bdge}, as used in previous time-dependent
studies of noncondensate dynamics of $\delta$-kicked BECs
\cite{gardiner_etal_pra_2000,zhang_etal_prl_2004}.  The GPE alone 
constitutes a zeroth-order description, although it may be possible to infer
higher-order processes from a pure GPE treatment \cite{monteiro_etal_prl_2009};
unlike GPE plus MBdGE, this is at least an internally consistent theoretical
description \cite{gardiner_morgan_pra_2007,blakie_etal_ap_2008}.


We take the $T=0$ equilibrium state (without driving) as our initial condition. 
The initial condensate mode is therefore spatially
homogeneous: $\psi = \sqrt{N_c/2\pi}$.  This sets $\lambda_0 = \is N_{c}/2\pi$,
and the initial stationary quasiparticle modes
\begin{equation}
\twovec{u_k(\x)}{v_k(\x)} = \frac{1}{2}\twovec{A_k + A_k^{-1}}{A_k - A_k^{-1}}\frac{e^{ik\x}}{\sqrt{2\pi}},
\label{eqn_init_qp}
\end{equation}
where $A_k = A_{-k} =  (1+4\lambda_0/k^2)^{-1/4}$. Hence, 
$N_{t} \equiv N - N_{c}= \int d\x \tilde{n}(\x,\x) =
(1/2) \sum_{k=1}^{\infty}(A_{k} - A_{k}^{-1})^{2}$, and we set $\lambda_{2} =
(\is/2\pi)[N - 1 + \sum_{k=1}^{\infty} (A_k^2 - 1)]$. To numerically determine
a self-consistent $T=0$ solution to \eqnsreft{eqn_ggpe}{eqn_bdge},
for given values of $N$ and $\is$, we set $N_c = N$, and then; (a) calculate $A_{k}$ up to a cut-off momentum $|k|= m$; (b)
determine $N_t$ from the $A_{k}$; (c) make the replacement $N_c = N-N_t$.
We repeat steps (a)--(c) until convergence.  To determine the driven
dynamics, we 
use a Fourier pseudospectral split-step method
\cite{Note4}; all simulations are converged in timestep, grid
size, and quasiparticle cut-off momentum $m$.

\begin{figure}
\includegraphics[width=\columnwidth]{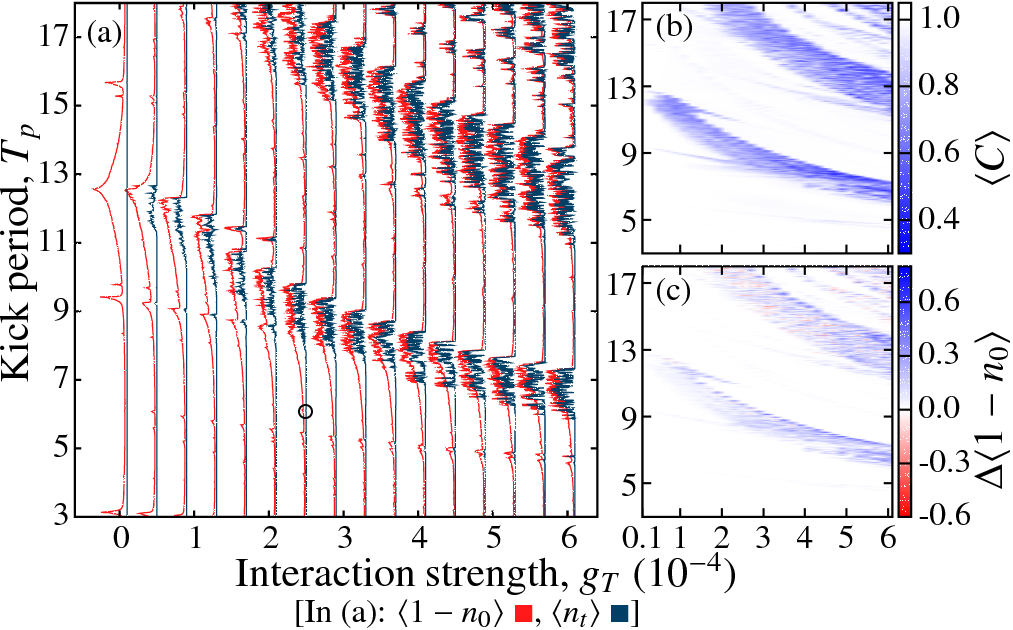}
\caption{(Color online) BEC response in the second-order description ($N=10^4$,
$\kappa=0.5$): (a) relative population of $k \ne 0$ momentum modes among all
atoms, $\avg{1-n_0}$, and noncondensate fraction, $\avg{n_t}$; (b) coherence
measure $\avg{C}$ \eqnrefp{eqn_g_def}; and (c) $\avg{1-n_0}$ as predicted by
the GPE, minus its value in the second-order description. Averages are taken
over the first 100 kicks.}
\label{figure_resonances_weak}
\end{figure}


In \figreft{fig_system_results}{(b)} we plot the condensate and noncondensate
fractions ($n_c = N_c/N$ and $n_t=N_t/N$) for parameters which, in the
first-order description, lead to rapid growth of the
noncondensate (becoming unphysical after $\sim 20$ kicks)
\figrefp{fig_system_results}{(b)(i)}. In the second-order description
\figrefp{fig_system_results}{(b)(ii)} the ``back-action'' of the noncondensate
rapidly damps out this growth, leading instead to
complementary oscillations in $n_t$ and $n_c$. We also track the overall
coherence of the system through
\begin{equation}
\g = \iint d\x d\x^\prime g_{1}(\x,\x^\prime) g_{1}(\x^\prime,\x),
\label{eqn_g_def}
\end{equation}
where $g_{1}(\x,\x^\prime) =
\avg{\hat{\Psi}^\dagger(\x^\prime)\hat{\Psi}(\x)}/N$ is the first-order
correlation function, and  compare the evolution of $\psi$ in the GGPE with
the GPE prediction ($\psi_{\rm GPE}$) through the fidelity
\begin{equation}
F = \frac{\left|\int d\x \psi^\ast_{\rm GPE}(\x) \psi(\x)\right|^2 }{NN_c}.
\label{eqn_f_def}
\end{equation}
The quantity $C$ equals unity only in the limit of a pure condensate, where the
noncondensate fraction is exactly zero (i.e., the single-body density matrix is
exactly factorizable). The GGPE then reduces to the GPE, and
$F=1$. Otherwise both $C$ and $F$ take values between zero and unity.   In
\figreft{fig_system_results}{(b)(iii)}  we observe that $C$ follows the
oscillations of $n_t$ closely, while $F$ shows larger
amplitude oscillations with revivals. 
Similar behavior persists across the $T_p$--$\is$ parameter space: in
\figreft{figure_resonances_weak}{(a)} we show the time averaged response to
weak driving ($\kappa = 0.5$), by plotting $1-n_{0}$ averaged
over the first 100 kicks;  here $n_k = \avg{\hat{a}_k^\dagger \hat{a}_k}/N$,
where $\hat{a}_k^\dagger$ creates an atom with momentum $k$. The structure of
this response over the range of \figreft{figure_resonances_weak}{(a)}, modeled
with the GPE alone, was recently elucidated \cite{monteiro_etal_prl_2009}:
the response is dominated by linear resonances corresponding to the first
two primary quantum resonances of the $\delta$-kicked rotor as $\is\rightarrow
0$ \cite{saunders_etal_pra_2007}. Higher-order linear resonances 
generally
decay with increasing $\is$, while nonlinear resonances appear, with no
analog in the linear regime \cite{monteiro_etal_prl_2009}. To 
first-order
(GPE plus MBdGE) all these resonant areas of parameter space are
associated with rapid growth of the noncondensate fraction $n_{t}$ due to
linear instabilities in the GPE dynamics  \cite{zhang_etal_prl_2004}. In
contrast, we find that in the second-order description (GGPE plus MBdGE) this
growth is damped out: throughout \figreft{figure_resonances_weak}{(a)} the
100-kick average of $n_t$ remains below 0.6. However, $\avg{n_t}$ is still
strongly enhanced in parameter regimes with significant resonant response
[large $\avg{1-n_{0}}$, as shown in \figreft{figure_resonances_weak}{(a)}]. The
predictions of the second-order description then differ considerably from the
standalone GPE description, as shown by the behavior of $\avg{\g}$
\figrefp{figure_resonances_weak}{(b)}, and the difference in response (as
measured by $\avg{1-n_{0}}$) between descriptions
\figrefp{figure_resonances_weak}{(c)}. Nonetheless, we find that the resonances
are located in the same regions of parameter space in both descriptions, and
that the asymmetric profiles and sharp cut-offs seen in
\cite{monteiro_etal_prl_2009} remain. 
Away from these resonances the GPE agrees well with the second-order
description, up to the 100 kicks we consider.

\begin{figure}
\includegraphics[width=\columnwidth]{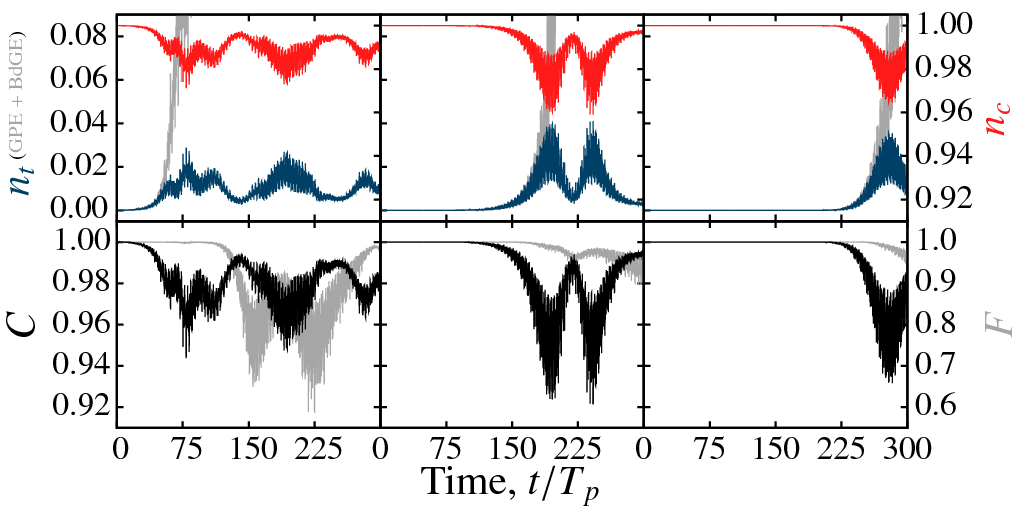}
\caption{(Color online) Comparison of first- and second-order descriptions
close to a nonlinear resonance explored in \cite{monteiro_etal_prl_2009} [$g_T
N = 2.5$, $T_p=6.12$, $\kappa=0.5$: circle in
\figreft{figure_resonances_weak}{(a)}].  Condensate and noncondensate
fractions $n_c$ and $n_t$, coherence measure $C$ \eqnrefp{eqn_g_def}, and the
fidelity of the condensate mode between descriptions, $F$ \eqnrefp{eqn_f_def},
are shown.  Columns correspond (left to right) to $N=10^4$, $10^8$, and
$10^{12}$; agreement of the initial growth in $n_t$ between the
second-order (dependent on $g_T$ and $N$) and first-order (dependent on $g_T
N$) descriptions over such a range is a useful test of the second-order
numerics.}
\label{figure_numbers_weak}
\end{figure}

\begin{figure}
\includegraphics[width=\columnwidth]{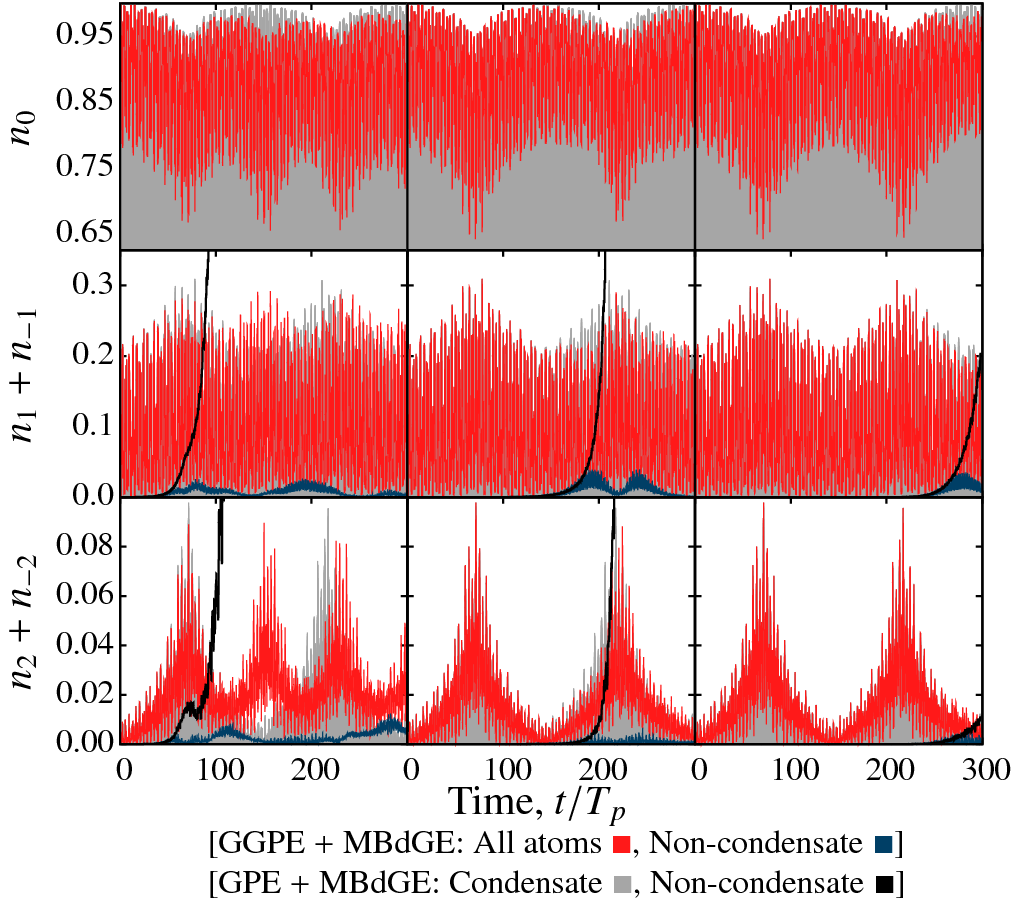}
\caption{(Color online) Relative populations, $n_k$, of low momentum modes in
the first- and second-order descriptions (parameters as in
\figreft{figure_numbers_weak}). In the second-order description $n_k$ is shown
among all atoms and among the noncondensate atoms; in the first-order
description $n_k$ is shown among condensate atoms (as a bar chart) and among the
noncondensate atoms (with a three-period moving average). Columns
correspond (left to right) to $N=10^4$, $10^8$, and $10^{12}$.}
\label{figure_kmodes_weak}
\end{figure}

In \figsreft{figure_numbers_weak}{}{figure_kmodes_weak}{} we compare the first-
and second-order descriptions, for varying $N$ but fixed $g_T N$, close to a
nonlinear resonance studied in \cite{monteiro_etal_prl_2009} [circle in
\figreft{figure_resonances_weak}{(a)}].  \figreftfull{figure_numbers_weak}{}
shows that the dynamics in the second-order description match the GPE for times
which increase with $N$. This increase is slow, however; for realistic $N$
($\ll10^{8}$) the loss of coherence, unaccounted for in the GPE and measured by
decay in $C$, quickly becomes significant.  Furthermore, in
\figreft{figure_kmodes_weak}{} we see that, on the same timescales associated
with significant decay in $C$, the dynamics of the relative populations
$n_{k}$, as studied in \cite{monteiro_etal_prl_2009} using the GPE, noticeably
differ in our second-order description. Compared to the first-order
description, rapid growth of the noncondensate begins at the same time in our
second-order description.  However, transfer of population to the noncondensate
is driven by, and sensitive to, atom-atom interactions. Hence, decreasing
population of the condensate, 
consistently accounted for in the second-order description, reduces the
mean-field interactions, and hence the rate of population transfer.  We observe
population oscillations between condensate and noncondensate fractions,
accompanied by oscillations in the coherence $C$ and fidelity $F$
\figrefp{figure_numbers_weak}{}. In \figreft{figure_kmodes_weak}{} we also
observe the exponential oscillations in $n_{2} + n_{-2}$ reported in
\cite{monteiro_etal_prl_2009}; however, for realistic atom numbers the
frequency of these oscillations is quickly increased by the presence of a
significant noncondensate fraction.

We have applied a fully dynamical, second-order, number-conserving approach
\cite{gardiner_morgan_pra_2007} to the $\delta$-kicked-rotor-BEC. In contrast
to a first-order approach, we observe that rapid growth of the noncondensate in
resonant parameter regimes is damped by our consistent treatment of the
condensate population and condensate-noncondensate interactions. Although our
description leads to different dynamics around resonant parameter regimes,
these regimes occur where the GPE predicts them.  Furthermore, our description
retains the cut-offs, but will typically strongly modify the exponential
oscillations predicted by the GPE. Extension of our second-order description to
other BEC systems offers exciting possibilities for future research.


We thank T. S. Monteiro and M. D. Lee for 
discussions, the UK
EPSRC (Grant No. EP/G056781/1), the Jack Dodd Centre (S.A.G.) and Durham
University (T.P.B.) for support.

\end{document}